\documentclass[letter,longauth]{aa} 
\usepackage{graphicx,amssymb}
\begin{document}
   \title{New constraints on the Mid-IR EBL 
from the HESS discovery of VHE $\gamma$-rays from 1ES\,0229+200}


\author{F. Aharonian\inst{1,13}
 \and A.G.~Akhperjanian \inst{2}
 \and U.~Barres de Almeida \inst{8} \thanks{supported by CAPES Foundation, Ministry of Education of Brazil}
 \and A.R.~Bazer-Bachi \inst{3}
 \and B.~Behera \inst{14}
 \and M.~Beilicke \inst{4}
 \and W.~Benbow \inst{1}
 \and K.~Bernl\"ohr \inst{1,5}
 \and C.~Boisson \inst{6}
 \and O.~Bolz \inst{1}
% ...
}

\offprints{Wystan.Benbow@mpi-hd.mpg.de or Luigi.Costamante@mpi-hd.mpg.de}

\institute{
Max-Planck-Institut f\"ur Kernphysik, Heidelberg, Germany
\and
 Yerevan Physics Institute, Armenia
\and
Centre d'Etude Spatiale des Rayonnements, CNRS/UPS, Toulouse, France
\and
Universit\"at Hamburg, Institut f\"ur Experimentalphysik, Germany
\and
Institut f\"ur Physik, Humboldt-Universit\"at zu Berlin, Germany
\and
LUTH, Observatoire de Paris, CNRS, Universit\'e Paris Diderot, Meudon, France
\and
DAPNIA/DSM/CEA, CE Saclay, Gif-sur-Yvette, France
\and
University of Durham, Department of Physics, U.K.
\and
Unit for Space Physics, North-West University, Potchefstroom, South Africa
\and
Laboratoire Leprince-Ringuet, Ecole Polytechnique, CNRS/IN2P3,
Palaiseau, France
\and 
Laboratoire d'Annecy-le-Vieux de Physique des Particules, CNRS/IN2P3,
Annecy-le-Vieux, France
\and
Astroparticule et Cosmologie (APC), CNRS, 
Universite Paris 7 Denis Diderot, France
\and
Dublin Institute for Advanced Studies, Ireland
\and
Landessternwarte, Universit\"at Heidelberg, K\"onigstuhl, Germany
% ...
}

   \date{Received 10 August 2007 / Accepted 23 September 2007}

 
  \abstract
   {}
   {To investigate the very high energy 
(VHE: $>$100 GeV) $\gamma$-ray emission from the 
high-frequency peaked BL\,Lac 1ES\,0229+200.}
   {Observations of 1ES\,0229+200 at energies above 580 GeV
were performed with the High Energy Stereoscopic System (HESS)
in 2005 and 2006.}
   {1ES\,0229+200 is discovered by HESS to be an emitter
of VHE photons.  A signal is detected at the 6.6$\sigma$ level in the 
HESS observations (41.8 h live time). The integral flux above 580 GeV
is $(9.4\pm1.5_{\rm stat}\pm1.9_{\rm syst}) \times 10^{-13}$
cm$^{-2}$\,s$^{-1}$, corresponding to $\sim$1.8\% of the 
flux observed from the Crab Nebula. The data show no evidence for 
significant variability on any time scale.  The observed spectrum is
characterized by a hard power law ($\Gamma = 2.50\pm0.19_{\rm stat}\pm0.10_{\rm syst}$) from 500 GeV to $\sim$15 TeV.}
   {The high-energy range and hardness of the observed spectrum,
   coupled with the object's relatively large redshift ($z=0.1396$),   
enable the strongest constraints so far on the density of the
Extragalactic Background Light (EBL) in the 
mid-infrared band. Assuming that the emitted spectrum 
is not harder than 
$\Gamma_{\rm int} \approx 1.5$, the HESS data
support an EBL spectrum $\propto \lambda^{-1}$ 
and density close to the lower limit from source counts measured by Spitzer,
confirming the previous indications from the HEGRA data
of 1ES\,1426+428 ($z=0.129$).
Irrespective of the EBL models used, the intrinsic spectrum 
of 1ES\,0229+200 is hard, thus locating 
the high-energy peak of its spectral energy distribution
above a few TeV.}


	\titlerunning{Discovery of VHE $\gamma$-rays from 1ES\,0229+200}
   \maketitle

\section{Introduction}

The active galactic nucleus (AGN) 
1ES\,0229+200 was initially discovered in the {\it Einstein} IPC Slew Survey
\cite{discovery_paper} and later identified as a BL Lac 
object (\cite{BL_id} 1993).  
It is now classified as a high-frequency peaked BL Lac (HBL) due to 
its X-ray-to-radio flux ratio \cite{classification}. 
The HBL is hosted 
by an elliptical galaxy with absolute magnitude
$M_R = -24.53$  \cite{elliptical_host},
located at a redshift of $z=0.1396$ \cite{BL_z}.
Based on its spectral energy distribution (SED) 1ES\,0229+200 is suggested
as a potential source of VHE $\gamma$-rays (\cite{stecker,luigi_AGN}).
However, despite several attempts, it has not been 
previously detected in the VHE regime.
The Whipple (\cite{Whipple_AGN_A,Whipple_AGN_B}), HEGRA \cite{HEGRA_AGN},
and Milagro \cite{Milagro_AGN} collaborations have each reported 
upper limits on the flux from 1ES\,0229+200 during various epochs.
The most constraining upper limit (99.9\% confidence level) 
on the flux is I($>$410 GeV) $< 2.76\times 10^{-12}$ cm$^{-2}$\,s$^{-1}$, 
based on $\sim$1 hour of HESS observations in 2004 \cite{HESS_AGN_UL}.  
The present discovery of VHE $\gamma$-rays from 1ES\,0229+200
makes it the most distant object detected at multi-TeV energies.
As a result its VHE spectrum can provide
important information on the EBL
(\cite{EBL_effect2}) in the mid-infrared band, where few measurements exist. 

\section{Observations and Analysis Technique}

1ES\,0229+200 was observed with the HESS array of
imaging atmospheric-Cherenkov telescopes \cite{HESS_jim}
for a total of 70.2 h (161 runs of $\sim$28 min each) in 2005 and 2006.  
After applying the standard HESS data-quality 
selection, 98 runs remain
yielding an exposure of 41.8 h live time at 
a mean zenith angle $Z_{\mathrm{mean}}=46^{\circ}$.
The data are calibrated as detailed in \cite{calib_paper}
and the standard HESS analysis tools \cite{std_analysis}
are used.  The event-selection criteria are performed
using the {\it standard cuts} \cite{std_analysis} resulting in a
post-analysis energy threshold of 580 GeV at $Z_{\mathrm{mean}}$.
A circular region of radius $\theta_{\mathrm{cut}}=0.11^{\circ}$ centered
on 1ES\,0229+200 is used for the on-source data. 
The background (off-source data) is estimated using
the {\it Reflected-Region} method \cite{bgmodel_paper}.
As the source direction was positioned $\pm$0.5$^{\circ}$ 
relative to the center of the field-of-view of the camera 
during the observations, this method allows the
simultaneous estimation of the background 
using the same data as the on-source measurement.
The significance of any excess is calculated following
the method of Equation (17) in \cite{lima}.
The energy of each event passing selection
is corrected for the absolute optical efficiency
of the system, using efficiencies determined from simulated and
observed muons \cite{HESS_crab}. This corrects any potential
long-term variations in the observed spectrum and flux 
due to the changing optical throughput of the HESS system. 
Results consistent with those presented
are also found using independent calibration and analysis chains.

\section{Results}

   \begin{table*}
      \caption{
        The MJD of the first and last night of HESS observations 
	of 1ES\,0229+200, 
        the live time of the observations, 
        the number of on- and off-source events measured, 
        the on/off normalization ($\alpha$),
        the excess, and the significance of the excess are given.
 	In addition, the integral flux above 580 GeV (assuming $\Gamma=2.50$), 
	and the corresponding percentage of the Crab Nebula flux
        above 580 GeV are shown.
	The $\chi^2$, degrees of freedom (NDF), 
	and $\chi^2$ probability, P($\chi^2$),
        for a fit of a constant to the flux binned
	by dark period within each year, 
	or yearly within the total, are also given.}
         \label{results}
        \centering
         \begin{tabular}{c c c c c c c c c c c c c}
            \hline\hline
            \noalign{\smallskip}
            Epoch & MJD & MJD & Time & On & Off & $\alpha$ & Excess & Sig & I($>$580 GeV) & Crab & $\chi^2$, NDF & P($\chi^2$)\\
             & First & Last & [h] & & & & &  [$\sigma$] & [10$^{-13}$\,cm$^{-2}$\,s$^{-1}$] & \% & & \\
            \noalign{\smallskip}
            \hline
            \noalign{\smallskip}
            2005 & 53614 & 53649 & 6.8 & 246 & 2238 & 0.09160 & 41 & 2.7 & 6.8$\pm$3.1$_{\rm stat}$$\pm$1.4$_{\rm syst}$ & 1.3 & 1.6\,,\,1 & 0.21\\
            2006 & 53967 & 54088 & 35.0 & 1344 & 12304 & 0.09136 & 220 & 6.1 & 10.0$\pm$1.7$_{\rm stat}$$\pm$2.0$_{\rm syst}$ & 1.9 & 1.5\,,\,3 & 0.68\\
            Total & 53614 & 54088 & 41.8 & 1590 & 14542 & 0.09140 & 261 & 6.6 & 9.4$\pm$1.5$_{\rm stat}$$\pm$1.9$_{\rm syst}$ & 1.8 & 0.8\,,\,1 & 0.37\\
            \noalign{\smallskip}
            \hline
       \end{tabular}
   \end{table*}

\begin{figure}
   \centering
      \includegraphics[width=0.43\textwidth]{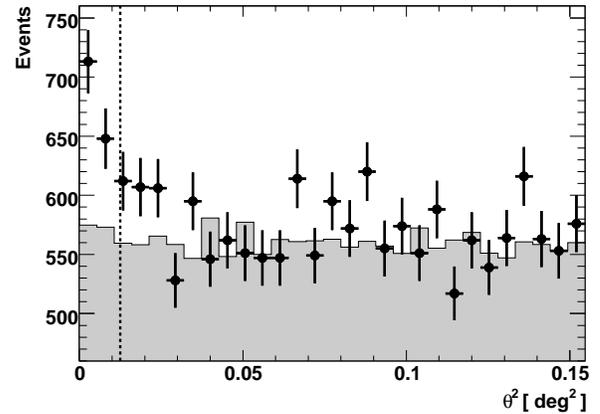} \\
      \caption{The distribution of $\theta^2$ for on-source 
        events (points) and
        normalized off-source events (shaded) from observations
        of 1ES\,0229+200.  The dashed line represents the cut 
        on $\theta^2$ applied to the data.}
         \label{thtsq_plot}
   \end{figure}

The results of the HESS observations of 1ES\,0229+200
in 2005 and 2006 are shown in Table~\ref{results}. A 
significant excess of 261 events (6.6$\sigma$)
from the direction of 1ES\,0229+200 is detected in the total data set.
The on-source and normalized off-source
distributions of the square of the angular difference between
the reconstructed shower position and
the source position ($\theta^{2}$)
are plotted in Figure~\ref{thtsq_plot} for all observations. 
The background is approximately flat in $\theta^{2}$ as expected, 
and there is a clear excess at small values of
$\theta^{2}$ corresponding to the observed signal.
A two-dimensional fit of the observed excess finds the
shape to be characteristic of a point source, located at
($\alpha_{\rm J2000}=2^{\rm h}32^{\rm m}53.2^{\rm s}\pm3.1^{\rm s}_{\rm stat}\pm1.3^{\rm s}_{\rm syst}$,
$\delta_{\rm J2000}=20^{\circ}16'21''\pm44''_{\rm stat}\pm20''_{\rm syst}$).
The excess, named HESS\,J0232+202,
is consistent with the position (\cite{BL_id} 1993) of the blazar
($\alpha_{\rm J2000}=2^{\rm h}32^{\rm m}48.6^{\rm s}$, 
$\delta_{\rm J2000}=20^{\circ}17'17''$) as expected, and
is therefore assumed to be associated with 1ES\,0229+200.

\begin{figure}
   \centering
      \includegraphics[width=0.43\textwidth]{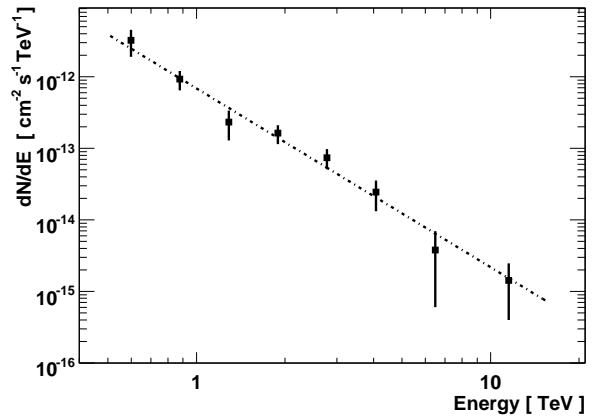} \\
      \caption{The VHE spectrum observed from 1ES\,0229+200. The
	line represents the best $\chi^2$ fit of a power law to
        the observed data. Only the statistical errors are shown.}
         \label{spectrum_plot}
   \end{figure}

Figure~\ref{spectrum_plot} shows the time-average photon spectrum
observed from 1ES\,0229+200.  The best $\chi^2$ fit of a power law 
(d$N$/d$E \sim E^{-\Gamma}$) to 
these data yields a photon index 
$\Gamma=2.50\pm0.19_{\rm stat}\pm0.10_{\rm syst}$, and a $\chi^2$ of 
3.6 for 6 degrees of freedom.  No evidence for significant features, 
such as a cutoff or break, are found in the measured spectrum.

Assuming the determined photon index of $\Gamma=2.50$,
the observed integral flux above 580 GeV is
I($>$580 GeV) $= (9.4\pm1.5_{\rm stat}\pm1.9_{\rm syst})
\times 10^{-13}$ cm$^{-2}$\,s$^{-1}$.
This corresponds to $\sim$1.8\% of I($>$580 GeV)
measured by HESS from the Crab Nebula \cite{HESS_crab}.
Table~\ref{results} shows the flux measured during each
year of HESS observations.  All values
are below the previously published VHE upper limits 
mentioned earlier.  The $\chi^2$ and degrees of freedom
for fits of a constant to the integral flux versus time
when binned by 
months within each year and years within the total, are also shown
in Table~\ref{results}.  As the $\chi^2$ probability for each fit,
as well as for fits of the flux binned nightly within each dark
period, is greater than 0.2, there is no evidence for variability 
on any time scale within the HESS data.

\section{Discussion}
\begin{figure}
   \centering
      \includegraphics[width=0.45\textwidth,height=6.5cm]{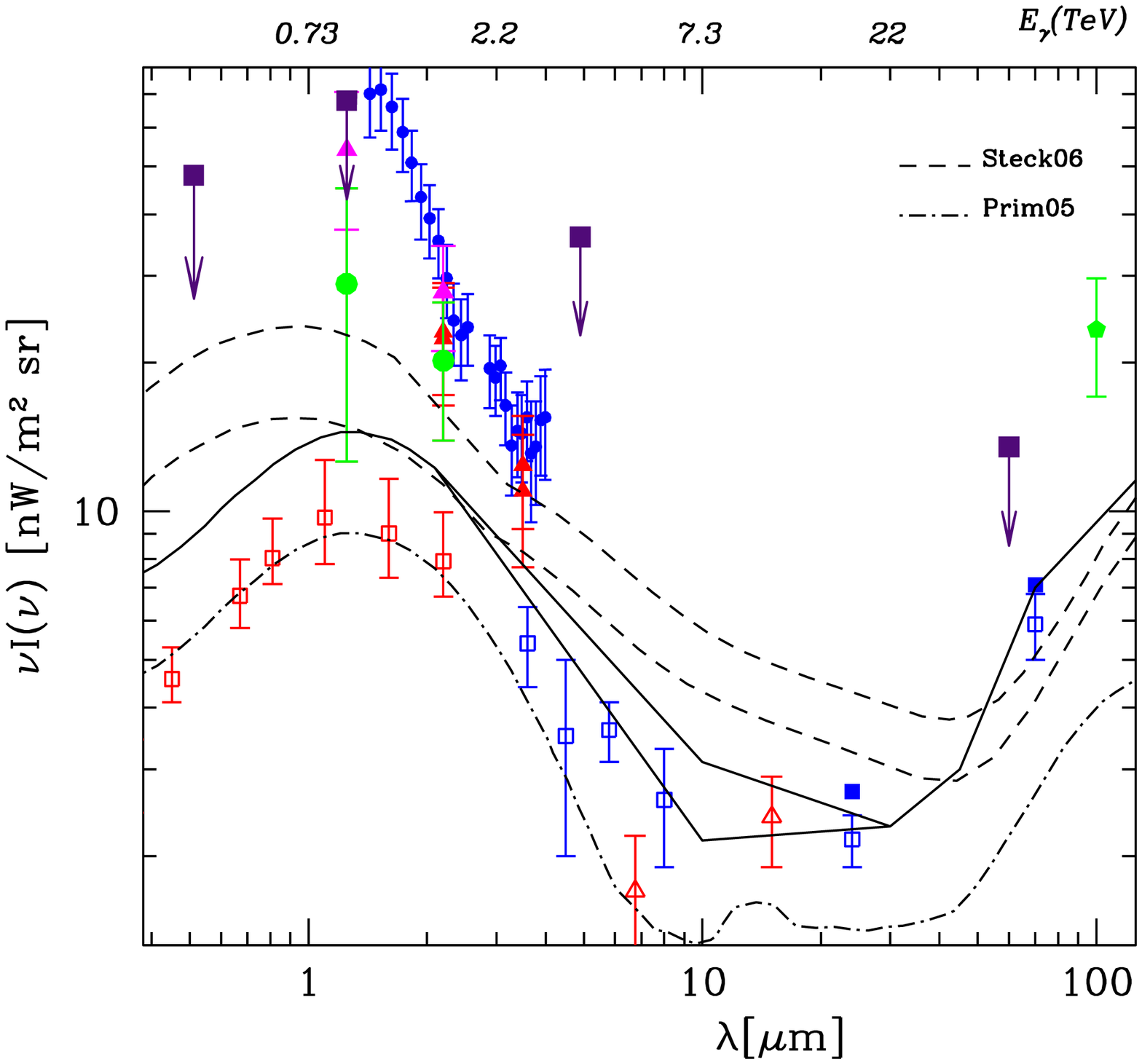} \\
      \caption{SED of the EBL 
	at wavelengths most relevant for the VHE spectrum of 
	1ES\,0229+200.
	The upper axis shows the energy scale corresponding to 
      	the peak of the $\gamma$-$\gamma$ cross-section. 
      	Data compilation from \cite{nature} (see references therein).
      	Open symbols: Lower limits given by the integrated light 
	from resolved sources.
      	From left to right: Hubble data (\cite{madaupozzetti}), 
	Spitzer data (\cite{fazio}), ISO data (triangles; \cite{altieri,elbaz}), 
	Spitzer data (\cite{dole}).
      	Dashed lines: {\it Fast evolution} (upper) and {\it baseline} (lower) models 
	of \cite{sms06}.
      	Dot-dashed line: Model ({\it Prim05}) of \cite{prim05}. 
	Solid lines: Scaled ($\times$1.6) {\it Prim05}, modified above
	2\,$\mu$m by a range of allowed MIR slopes, 
	i.e. those which yield an intrinsic spectrum $\Gamma_{\rm int}=1.5$ and 
	$1.15 \approx 1.5-0.23_{\rm stat}-0.1_{\rm syst}$.
	The fluxes at 10\,$\mu$m are 2.15 and 3.1 nW m$^{-2}$ sr$^{-1}$, 
	respectively. }
         \label{ebl}
   \end{figure}

VHE photons from extragalactic sources are expected to suffer 
absorption from interactions
($\gamma_{_{\mathrm{VHE}}}$\,$\gamma_{_{\mathrm{EBL}}}$\,$\rightarrow$\,$e^{+}$\,$e^{-}$; \cite{EBL_abs})
with EBL photons (see, e.g., the review of \cite{felix}).
The energy-dependent opacity $\tau(z,E)$ causes a 
deformation of the observed VHE spectrum 
($F_{\mathrm{obs}}(E) = F_{\mathrm{int}}(E)$\,e$^{-\tau(z,E)}$)
which depends on the EBL SED (Figure~\ref{ebl}).
With assumptions about a source's intrinsic VHE spectrum,
limits on the EBL density can be derived.
Recently, the hard 0.2$-$2 TeV spectra observed from
1ES\,1101$-$232 and H\,2356$-$309
constrained the EBL around 1$-$2\,$\mu$m to a significantly low level, 
close to the lower limits given by the integrated light 
from resolved galaxies (\cite{nature,gerd_1101}).

Apart from 1ES\,1426+428 (\cite{hegra1426}),
1ES\,0229+200 is currently the only VHE source at $z$$>$0.1 
whose spectrum has been measured up to 10 TeV.
Since the peak of the $\gamma$-$\gamma$ cross-section occurs at
$\lambda^\ast\approx1.4\;(E_{\gamma}/\,1 \, {\rm TeV})\, \mu{\rm m}$,
these VHE spectra probe the EBL 
in the near (NIR) to mid-infrared (MIR) range ($\approx$ 2$-$20\,$\mu$m),
where direct estimates are missing 
due to increasing foreground emission from interplanetary dust
and where EBL models differ significantly in the slope of the SED
(Figure~\ref{ebl}). 
This difference is exemplified by comparison of the models by Primack et al. 
(2005, hereafter {\it Prim05}) and 
Stecker et al. (2006, hereafter {\it Steck06}).
Aside from normalization, the NIR-MIR spectrum 
in the {\it Prim05} model, as well as in the models by \cite{kneiske},
follows the characteristic
decline of the old stellar component in the SED of galaxies, 
which behaves approximately as $\nu I(\nu)\propto \lambda^{-1}$
at these wavelengths.
The corresponding photon number density, 
$n(\epsilon) \propto \epsilon^{-1}$,
makes the optical depth almost constant with energy 
between one and several TeV (\cite{felix}).  This
causes a characteristic flattening feature in the attenuation curves.
This flattening will be
imprinted in the observed VHE spectra 
for an intrinsic power-law spectrum.
In contrast, the EBL decline is much flatter than $\lambda^{-1}$
for both the {\it baseline} and {\it fast evolution} 
versions of the {\it Steck06} model.  This yields a 
rapidly increasing opacity above 1 TeV 
which should translate into a strong and continuous steepening of
the observed VHE spectra (unless counter-balanced by 
upturns in the intrinsic spectrum).  It should be noted
that, regardless of the EBL model, the optical depth is
large (i.e. $\tau$\,$>$\,1) for all energies relevant here. Therefore,
the observed spectrum of 1ES\,0229+200 is very 
sensitive to $\tau$$(E)$ and hence to the shape of the EBL SED.

The intrinsic spectrum of 1ES\,0229+200 can be 
reconstructed\footnote{Throughout the paper, a cosmology with 
H$_0$=$70\;\rm km/s/Mpc$, $\Omega_m$=$0.3$ and $\Omega_{\Lambda}$=$0.7$
is used for the calculation of $\tau(z,E)$.}
by multiplying the individual data points by e$^{\tau(z,E)}$.
For the two {\it Steck06} models\footnote{The optical depths were 
calculated by integrating over the respective EBL SED. 
The methodology was checked with the 
prescription in \cite{sms06} for the nearest 
tabulated redshift (z=0.117), obtaining consistent
values ($\Delta\Gamma<$0.05) for the intrinsic spectrum.}, 
the 1ES\,0229+200 absorption-corrected spectra are well 
fit by power-law functions (d$N$/d$E \sim E^{-\Gamma_{\rm int}}$) 
with hard photon 
indices $\Gamma_{\rm int}=0.6\pm0.3$ ({\it baseline} EBL) and 
$\Gamma_{\rm int}=0.1\pm0.3$ ({\it fast evolution} EBL).  
Such hard spectra are not easily explained by commonly used 
leptonic or hadronic scenarios for the $\gamma$-ray emission 
of BL\,Lac objects, unless invoking extreme assumptions 
like mono-energetic particle distributions, 
either in the shocked plasma (see, e.g., \cite{katarzynski})
or as a cold wind with very high ($\sim$$10^7$) Lorentz factors 
(\cite{bulkmotion}). Further, a specific dependence 
of source parameters on redshift 
would be required, in order to explain the absence of these 
hard features in all nearby ($z$$<$0.1), 
less-absorbed VHE-bright HBL.

The {\it Prim05} model yields $\Gamma_{\rm int}=1.92\pm0.22$. 
Since this model falls below the recent lower limits given 
by Spitzer source counts at 
MIR (\cite{fazio}; see Figure~\ref{ebl}) wavelengths,  and thus 
underestimates the attenuation at several TeV, modified EBL 
shapes consistent with the lower limits from source
counts are considered in the following.

In standard blazar models, VHE $\gamma$-rays are produced by 
particles accelerated in shocks.  For a wide range of 
conditions the resulting photon spectrum is expected
to have $\Gamma_{\rm int}\gtrsim1.5$ (\cite{nature}).  
This can be used to constrain the EBL.
It should be noted that \cite{stecker_hard} have demonstrated 
that intrinsic spectra as hard as $\Gamma_{\rm int}\approx1.0$ 
are possible in relativistic shock acceleration scenarios. 
In such situations, the derived EBL SED constraints will be relaxed.
Following Aharonian et al. (2006a),
the intrinsic photon spectrum is 
reconstructed from an EBL SED shape\footnote{In the 
calculation of $\tau$ for these test-SEDs,
the effects of galaxy evolution were approximated by modifying
the cosmological dependence of the photon number density
as $(1+z)^{3-k}$. With k=1.2, the values of $\tau(z)$ are in 
good agreement with those from the {\it Prim05} model at $z<1$. 
Regardless, evolution effects are small at $z=0.140$. 
The difference between including or not this effect
is $\Delta\Gamma<0.08$.} and then tested for 
compatibility with $\Gamma_{\rm int}\gtrsim1.5$.
Different EBL shapes can be derived for examination, either by 
scaling a fixed EBL shape or by changing the NIR/MIR flux ratio (thus
testing different EBL slopes). 

In the observed energy range of 1ES\,0229+200, 
a simple scaling of EBL SED shapes $\approx\lambda^{-1}$, like the
P1.0 curve in Aharonian et al. 2006a or the {\it Prim05} shape,
does not strongly affect the reconstructed intrinsic spectrum
due to the weak energy dependence of the optical depth.
Indeed, the spectrum is only modified due to the difference in $\tau$ for
low and high energy $\gamma$-rays, which is not the case 
here where the EBL mainly causes attenuation.
The limit for the overall scaling factor from the HESS 1ES\,0229+200 data is
0.8$\times$ P1.0 (i.e. P0.8), or 1.8$\times$ {\it Prim05}.
The limit on the flux is above those obtained from other VHE sources
(\cite{nature}).

Different EBL spectral slopes are tested using an EBL template
constructed from a fixed scaling of the {\it Prim05} shape up to 2$\mu$m
and an interpolation above this wavelength.
The MIR-EBL slope is varied using a linear (in log-log scale)
interpolation between 2, 10, and 30$\mu$m,
where only the 10\,$\mu$m flux is changed (see Figure~\ref{ebl}).
The scaled EBL level is chosen to be 1.6$\times${\it Prim05}, 
which corresponds to the EBL upper limit at 1$-$2\,$\mu$m as 
derived by Aharonian et al. (2006a).  This NIR EBL level allows
for the highest possible MIR fluxes, since a
high flux at 1$-$2\,$\mu$m softens the intrinsic
spectrum of 1ES\,0229+200
in the energy band measured here by HESS.  Thus, much higher
overall EBL levels are compatible with $\Gamma_{\rm int}>1.5$.
For a flux at 10\,$\mu$m of  
2.15 nW m$^{-2}$ sr$^{-1}$ the intrinsic
spectrum is $\Gamma_{\rm int}=1.5$. 
A limiting 10$\mu$m flux of 3.1 nW m$^{-2}$ sr$^{-1}$ is derived
by taking into account the errors on the photon index (i.e. where 
$\Gamma_{\rm int}=1.15 \approx 1.5-0.23_{\rm stat}-0.1_{\rm syst}$).
The EBL SED between 2 and 10\,$\mu$m ($\propto \lambda^{-\alpha}$) is thus
constrained to a slope $\alpha\gtrsim1.1\pm0.25$.
A $\Gamma_{\rm int}\sim$1.5 spectrum is also obtained with an EBL 
SED at the level of the source counts, connecting the optical data
points with the Spitzer data at 8\,$\mu$m (for an $\alpha\sim0.9$).
Clearly the HESS data do not support flatter EBL shapes in the MIR,
confirming a trend already suggested by 
the HEGRA data on 1ES\,1426+428 (\cite{hegra1426}).

  \begin{figure}
   \centering
      \includegraphics[width=0.45\textwidth]{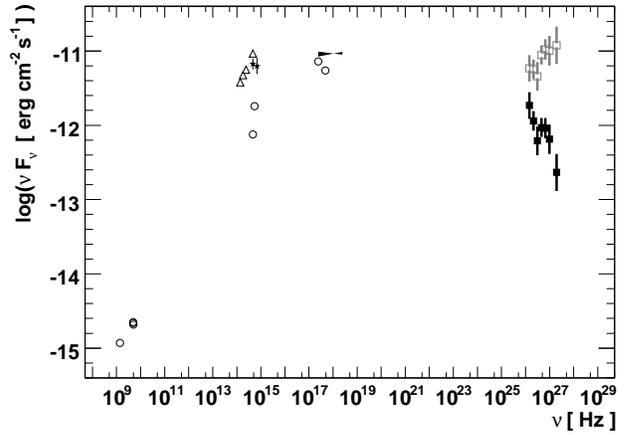} \\
      \caption{The broad-band SED of 1ES\,0229+200. 
      The ATOM optical measurements (stars) are simultaneous
      to some of the 2006 HESS data. All other data 
      are non-simultaneous.
      Both the observed and intrinsic (i.e. de-absorbed 
      with a 1.6$\times$ {\it Prim05} EBL shape, yielding 
      $\Gamma_{\rm int}=1.58\pm0.22$)
      VHE spectra are shown, each with the two highest-energy HESS points
      combined for visibility purposes.
      Using a lower EBL density results in a softer intrinsic spectrum, 
      but even the very low density of {\it Prim05} 
      yields $\Gamma_{\rm int} < 2.0$.
      The X-ray data are 
      from {\it Einstein} (2 keV), ROSAT (1 keV; \cite{ROSAT_obs}),
      and BeppoSAX (butterfly; \cite{SAX_obs}) observations.
      The radio, NIR (2MASS) and optical data are from the NED archive, 
      \cite{HST},
      \cite{Perlman_MWL}, and \cite{BL_id} (1993).
      The ATOM fluxes and the
      data marked with open triangles do not have the contribution
      of the host galaxy subtracted.
}
         \label{new_SED}
   \end{figure}

\section{Conclusions} 
The HESS spectrum of 1ES\,0229+200 leads to two important conclusions.
First, unless the intrinsic BL\,Lac spectrum is considerably 
harder\footnote{See, e.g., \cite{stecker_hard} for scenarios that demonstrate 
$\Gamma_{int}$ as hard as 1.0 is possible in shock acceleration models.}
than 1.5, the EBL density in the MIR range 
follows a spectrum $\propto \lambda^{-1}$
and is very likely close to the lower limits
given by galaxy counts as determined by Spitzer and ISO data. 
This implies that the sources resolved by Spitzer 
seem to account for the bulk of the diffuse 5$-$10\,$\mu$m background. 
Second, almost independently of the details of the EBL SED,
the intrinsic spectrum of  1ES\,0229+200 is significantly hard 
up to $\sim$10 TeV. Thus the high-energy peak of its SED 
(shown in Figure~\ref{new_SED}) is located at energies above a few TeV,
among the highest ever seen from a blazar. The extension of the SED 
to VHE energies provides clear evidence that non-thermal
processes are at work in 1ES0229+200.  The SED also includes
optical (Bessel B and R filters) measurements of 1ES\,0229+200, 
$m_{\rm R} = 16.2$ and $m_{\rm B} = 17.2$,
conducted with the ATOM telescope \cite{ATOM_ref} located at the HESS site.
These data are simultaneous with some of the 2006 HESS observations.
The ATOM fluxes are constant in time and fall below archival 
values.  Although only limited evidence for a VHE-optical flux 
correlation exists \cite{Mkn180}, 
the low optical state might suggest a correspondingly low VHE state.
These results strongly motivate further VHE observations of 1ES\,0229+200,
as well as contemporaneous observations at lower energies to enable 
SED modeling.  

\begin{acknowledgements}

The support of the Namibian authorities and of the University of Namibia
in facilitating the construction and operation of H.E.S.S. is gratefully
acknowledged, as is the support by the German Ministry for Education and
Research (BMBF), the Max Planck Society, the French Ministry for Research,
the CNRS-IN2P3 and the Astroparticle Interdisciplinary Programme of the
CNRS, the U.K. Science and Technology Facilities Council (STFC),
the IPNP of the Charles University, the Polish Ministry of Science and 
Higher Education, the South African Department of
Science and Technology and National Research Foundation, and by the
University of Namibia. We appreciate the excellent work of the technical
support staff in Berlin, Durham, Hamburg, Heidelberg, Palaiseau, Paris,
Saclay, and in Namibia in the construction and operation of the
equipment.  This research has made use of the NASA/IPAC Extragalactic 
Database (NED) which is operated by the Jet Propulsion Laboratory, 
California Institute of Technology, 
under contract with the National Aeronautics and Space Administration.

\end{acknowledgements}


\begin{thebibliography}{}

 \bibitem[Aharonian 2001]{felix}
        Aharonian, F. 2001, Proceedings 27th ICRC (Hamburg), 	
	Invited, Rapporteur, and Highlight Papers, 250

 \bibitem[Aharonian et al. 2002]{bulkmotion}
        Aharonian, F., Timokhin, A. \& Plyasheshnikov, A.V. 2002, A\&A, 384, 834

 \bibitem[Aharonian et al. 2003]{hegra1426}
        Aharonian, F., et al. (HEGRA Collaboration) 2003, A\&A, 403, 523

 \bibitem[(Aharonian et al. 2004a)]{HEGRA_AGN}
        Aharonian, F., et al. (HEGRA Collaboration) 2004a, A\&A, 421, 529

  \bibitem[Aharonian et al. (2004b)]{calib_paper}
        Aharonian, F., et al. (HESS Collaboration) 2004b, 
	Astroparticle Physics, 22, 109

 \bibitem[(Aharonian et al. 2005)]{HESS_AGN_UL}
        Aharonian, F., et al. (HESS Collaboration) 2005, A\&A, 441, 467

 \bibitem[Aharonian et al. 2006a]{nature}
        Aharonian, F., et al. (HESS Collaboration) 2006a, Nature, 440, 1018

 \bibitem[(Aharonian et al. 2006b)]{HESS_crab}
        Aharonian, F., et al. (HESS Collaboration) 2006b, A\&A, 457, 899

 \bibitem[Aharonian et al. 2007]{gerd_1101}
        Aharonian, F., et al. (HESS Collaboration) 2007, A\&A, 470, 475
	
  \bibitem[(Albert et al. 2006)]{Mkn180}
        Albert, J., Aliu, E., Anderhub, H., et al. 2006, ApJ, 648, L105

  \bibitem[Altieri et al. 1999]{altieri}
        Altieri, B., Metcalfe, L. , Kneib, J.P., et al. 1999, A\&A, 343, L65
        
 \bibitem[(Benbow 2005)]{std_analysis} 
	Benbow, W. 2005, Proceedings of Towards a Network of 
        Atmospheric Cherenkov Detectors VII (Palaiseau), 163

 \bibitem[(Berge et al. 2007)]{bgmodel_paper} 
	Berge, D., Funk, S. \& Hinton, J. 2007, 
	A\&A, 466, 1219
 
 \bibitem[Brinkmann et al. 1995]{ROSAT_obs}
	Brinkmann, W., Siebert, J., Reich, W., et al. 1995, 
	A\&AS, 109, 147

 \bibitem[Costamante \& Ghisellini 2002]{luigi_AGN}
        Costamante, L. \& Ghisellini G. 2002, A\&A, 384, 56

 \bibitem[de la Calle Perez et al. 2003]{Whipple_AGN_A}
         de la Calle Perez, I., Bond, I.H., Boyle, P.J., et al. 2003, 
	ApJ, 599, 909

 \bibitem[Dole et al. 2006]{dole}  
         Dole, H., Lagache, G., Puget, J.-L., et al. 2006, A\&A, 451, 417 
	 
 \bibitem[Donato et al. 2005]{SAX_obs} 
	Donato, D., Sambruna, R.M. \&  Gliozzi, M. 2005, A\&A, 433, 1163

 \bibitem[Elbaz et al. 2002]{elbaz} 
        Elbaz, D., Cesarsky, C.J., Chanial, P., et al. 2002, A\&A 384, 848

 \bibitem[(Elvis et al. 1992)]{discovery_paper} 
	Elvis, M., Plummer, D., Schachter, J., \& Fabbiano, G. 1992, 
	ApJS, 80, 257

 \bibitem[(Falomo \& Kotilainen 1999)]{elliptical_host} 
	Falomo, R. \& Kotilainen, J.K. 1999, A\&A, 352, 85

 \bibitem[Fazio et al. 2004]{fazio} 
        Fazio, G.G., Ashby, M.L.N., Barmby, P., et al. 2004, ApJS, 154, 39

 \bibitem[(Giommi et al. 1995)]{classification} 
	Giommi, P., Ansari, S.G., \& Micol, A. 1995, A\&AS, 109, 267

\bibitem[Gould \& Schr\'eder 1967]{EBL_abs}
	Gould, R.J. \& Schr\'eder, G.P. 1967, Physical Review, 155, 1408

 \bibitem[Hauser \& Dwek 2001]{EBL_effect2}
        Hauser, M.G. \& Dwek, E. 2001, ARA\&A, 39, 249

 \bibitem[(Hauser et al. 2004)]{ATOM_ref}
        Hauser, M., M\"ollenhoff, C., P\"uhlhofer, G., et al. 2004, 
	Astronomische Nachrichten, 325, 659

 \bibitem[(Hinton 2004)]{HESS_jim} 
	Hinton, J. 2004, New Astron Rev, 48, 331

 \bibitem[Horan et al. 2004]{Whipple_AGN_B} 
	Horan, D., Badran, H.M., Bond, I.H., et al. 2004, ApJ, 603, 51

 \bibitem[Katarzynski et al. 2006]{katarzynski} 
	Katarzynski, K., Ghisellini, G., Tavecchio, F., et al. 2006, MNRAS, 
	368, L52

 \bibitem[Kneiske et al. (2004)]{kneiske}
	Kneiske, T.M., Bretz, T., Mannheim, K. \& Hartmann, D.H. 2004,
	A\&A, 413, 807

 \bibitem[Li \& Ma (1983)]{lima} 
	Li, T. \& Ma, Y. 1983, ApJ, 272, 317

 \bibitem[Madau \& Pozzetti 2000]{madaupozzetti} 
        Madau, P. \& Pozzetti, L. 2000, MNRAS, 312, L9

 \bibitem[Perlman et al. (1996)]{Perlman_MWL}
	Perlman, E.S., Stocke, J.T., Schachter, J.F., et al. 1996, 
	ApJS, 104, 251

 \bibitem[Primack et al. (2005)]{prim05}
       Primack, J.R., Bullock, J.S. \& Somerville, R.S. 2005, 
	AIP Conference Proceedings, 745, 23

 \bibitem[Schachter et al.]{BL_id} 
	Schachter, J.F., Stocke, J.T., Perlman, E., et al. 1993, ApJ, 412, 541

 \bibitem[Stecker, de Jager \& Salamon 1996]{stecker}
         Stecker, F.W., de Jager, O.C., \& Salamon, M.H. 1996, ApJ, 473, L75

 \bibitem[Stecker et al. (2006)]{sms06}
         Stecker, F.W., Malkan, M.A., \& Scully S.T. 2006, ApJ, 648, 774, with Erratum

 \bibitem[Stecker et al. (2007)]{stecker_hard}
	Stecker, F.W., Baring, M.G., \& Summerlin, E.J. 2007, ApJ, 667, L29

 \bibitem[Urry et al. (2000)]{HST}
	Urry, C.M., Scarpa, R., O'Dowd, M., Falomo, R., et al. 2000, 
	ApJ, 532, 816

 \bibitem[(Williams 2005)]{Milagro_AGN} 
	Williams, D.A. 2005, AIP Conference Proceedings, 745, 499

 \bibitem[(Woo et al. 2005)]{BL_z} 
	Woo, J.H., Urry, C.M., Van der Marel, R.P., et al. 2005, ApJ, 631, 762

\end{thebibliography}
\end{document}